\documentclass[12pt,aps]{revtex4}

\begin{document}
\title{The limiting nuclear polarization in a  quantum dot under
optical electron-spin orientation and applicability of the box-model
of the electron-nuclear dynamics.}
\author{Kozlov G.G.}
\hskip100pt {\it e}-mail:  gkozlov@photonics.phys.spbu.ru

\begin{abstract}
For the model Hamiltonian describing the electron-nuclear dynamics
of a quantum dot, we obtained an exact expression for  the
limiting nuclear polarization as a function of the number of
groups of equivalent nuclei.  It is shown that the refinement of
the model Hamiltonian by increasing the number of the groups
 results in a slow growth of the limiting nuclear polarization.
This allowed us to put forward arguments in favor of applicability
of the box-model (with all the nuclei being equivalent)  for
description of the electron-nuclear spin dynamics within the time
intervals of around hundreds of periods of the optical
orientation.
\end{abstract}
\maketitle

\section{Introduction}

 Spin systems involving electron spin coupled by contact interaction with a
 large number of nuclear spins are well known in radio-spectroscopy  (rf spectroscopy). The
 appropriate quantum-mechanical models are well developed for the case of large
 magnetic fields typical for the rf spectroscopy (when the Zeeman energy of an
 electron spin in an external magnetic field substantially exceeds the energy of
 electron-nuclear interaction). In recent years, there have appeared new objects
 -- quantum dots, whose electron-nuclear dynamics can be described by similar
 models \cite{Sch}.

 A typical experiment on spin dynamics of a quantum dot implies
 observation of time dependence of spin of an electron localized in a quantum
 dot initially spin-oriented by a short polarized laser pulse. Observation of
 the electron spin dynamics, in these experiments, is performed within the time
 intervals when the contact interaction of this spin with nuclear spins of the
 quantum dot appears to be predominant. The experiments are usually interpreted
 using the Hamiltonian

\begin{equation}
H=\omega S_z+\sum_i \bigg\{A_{i\|}S_zI_{zi}+A_{i\bot}[S^+I_i^-+S^-I_i^+]\bigg\}
\end{equation}

where $\omega$ is the external magnetic field in frequency units, and
 $A_{i\|}$ and $A_{i\bot}$ -- are
 tensor components of the hyperfine interaction between the i-th nuclear and the
electron spin,
  $S_z$  ($I_{iz}$) --  are the operators of the z-projection of the electron spin
(i-th nuclear spin), and
$S^\pm$  ($I_i^\pm$) î
are the electron (nuclear) spin z-projection rising and
lowering operators. In contrast to the radio-spectroscopic experiments, these
experiments are carried out in a wide range of magnetic fields. For this reason,
theoretical methods of rf spectroscopy cannot be used to analyze experiments
with quantum dots. A great number of nuclei interacting with the electron spin
makes it impossible to solve appropriate problems numerically using computers.
This is why exactly solvable models of the electron-nuclear spin dynamics are
of particular interest even when the exact solution is obtained at the expense
of assumptions whose plausibility cannot be reliably evaluated.

In the theory of spin dynamics of the quantum dots, an exact
solution can be obtained in the framework of the box-model
\cite{Sem,Zha,Koz},
  which implies that the electron spin is coupled in
the same way  with all the nuclear spins of the quantum dot. Thus, the box-model
postulates that the electron density (electron wavefunction module squared) is
constant within the quantum dot. The calculations of the electron-nuclear spin
dynamics fulfilled in the framework of the box-model have shown its high
efficiency. With the aid of the box-model, it became possible: (i) to describe
electron spin dynamics after optical orientation,
  (ii) to obtain the
magnetic-field dependence of the residual electron polarization of the quantum
dot,
 (iii) to show that, in the experiments on spin dynamics of quantum dots,
one should expect a strong deviation of the nuclear state from equilibrium,
 (iv)
to describe the effect of replicas arising in the electron spin dynamics (the
effect results from appearance
  of regularity in the nuclear density matrix under
condition of periodic optical orientation of the electron spin), and (v) to
predict the echo-effect in spin dynamics of a {\it single} quantum dot after a
 $\pi$-pulse
of the magnetic field. These facts show that the simplifications
laid down into the basis of the box-model are not important for
description of a number of experiments.

The box-model, however, has at least one property that restricts
its application and needs to be additionally discussed. The point
is that, in the framework of the box-model, it is impossible to
obtain a high nuclear polarization for the constant-sign
orientation of the
 electron spin. For quantum
dots with $10^4$ nuclei, the limiting polarization calculated in
the framework of the box-model lies in the range of 1$\%$, whereas
there have been reported experimental observations of the nuclear
polarization of several tens of percents \cite{Gam,Forty,Fifty}.
  The above properties of the box-model are related to the fact that all
nuclei of the quantum dot, in this model, are considered to be equivalent (with
respect to interaction with the electron spin localized in the quantum dot) and
the Hamiltonian of the electron-nuclear
 interaction depends on the {\it total} nuclear
moment. In real systems, the electron density is not constant over the quantum
dot. This fact violates the restriction imposed on the limiting nuclear polarization
following from the box-model.

The goal of the present work  is to study dependence of the
limiting nuclear polarization arising in the quantum dot upon
optical orientation of the electron spin on the degree of
nonuniformity of the electron density in the quantum dot. The
spatial nonuniformity of electron density in the quantum dot can
be taken into account by characterizing this nonuniformity by
surfaces of equal density and assuming that the nuclei located
between the neighboring surfaces are equivalent with respect to
 their coupling to the electron
spin in the quantum dot. To be more precise, we will choose, in the range of
variation of the electron density, $n$ levels
  $|\Psi_i|^2, i=1,...,n$, so that: $0=|\Psi_0|^2<|\Psi_1|^2<
 |\Psi_2|^2<...<|\Psi_n|^2\equiv |\Psi|^2_{max}$.
   The nuclei of the region
with electron density  $|\Psi|^2$ lying in the range
$|\Psi_{i-1}|^2<|\Psi|^2<|\Psi_{i}^2|$
  are referred to i-th group and
are considered to be equally coupled to the electron. In this way, we obtain $n$
groups of equivalent nuclei. The accuracy of such a description will improve
 with increasing number of the groups. The box-model evidently corresponds
to the case of $n = 1$. The main result of this paper is the exact
expression for the limiting nuclear polarization of the
electron-nuclear system with a given number $n$ of the groups of
equivalent nuclei
 and with a given number  of nuclei in
each group. For large number of nuclei in the groups, the limiting
polarization appears to be higher than that in the box-model
  by a factor of $\sqrt{n}$ as the upper limit.

\section{The limiting nuclear polarization for a given number of groups of equivalent
nuclei }

In what follows, we will consider the electron-nuclear spin system with
the elementary nuclear angular momenta equal to $1/2$.
  Recall that  in this case
the expression for the limiting nuclear polarization $P_1$
  for the box-model with
the number of equivalent nuclei equal to $2N$, has the form
\cite{Koz}:

 \begin{equation}
P_1=P(N)\equiv{1\over 2N}{1\over 2^{2N}}\sum_{J=0}^N \Gamma_N(J)(2J+1)J\approx{1\over\sqrt{N\pi}}
 \end{equation}
$$
\Gamma_N(J)=C_{2N}^{N-J}-C_{2N}^{N-J-1}, \hskip5mm J=0,1,...,N
$$

 The approximate equality is valid for large $N$.

 Now, we will extend the
 box-model in the following way. We will assume that the spin system has two,
 rather than one, group of the equivalent nuclei characterized by two tensor
 constants of contact interaction ${\bf A}_1$ and ${\bf A}_2$
  and containing, respectively, $2N_1$
 and $2N_2$ nuclear spins.
 The relevant Hamiltonian written in frequency units will
 be given by:

\begin{equation}
  {\cal H}_2=\omega S_z+H_1+H_2
  \end{equation}

where the first term corresponds to Zeeman energy
of the electron spin in the external magnetic field $\omega$,
  and the operators

\begin{equation}
H_i\equiv A_{i\|}S_z I_{iz}+A_{i\bot}(I_i^+S^-+I_i^-S^+), \hskip5mm i=1,2
\end{equation}

describe contact interaction of the electron spin with the nuclear spins of the
first and second groups of equivalent nuclei.  Here ${\bf S}$ and ${\bf I}_i$
 are operators of electron angular momentum and total angular momentum of i-th nuclear group.

  Let us introduce the
repreasentation of the functions $|S,J_1,L_1,J_2,L_2,\alpha\rangle$,
  where $S = \pm 1/2$ is the quantum number of
the electron spin z-projection,
 $J_i$ and $L_i$ --
  are the quantum numbers of the total
angular momentum of the i-th nuclear group and its z-projection,
  respectively, $\alpha$
is the set of all other quantum numbers needed for the state to be uniquely
specified. In this case, the total angular momentum squares of the i-th nuclear
group is
 $I^2_i=J_i(J_i+1)$
(in the units $\hbar^2$) and the possible values of
  $L_i$ are:
$L_i=-J_i,1-J_i,...,J_i-1,J_i$.
  As has been
shown in \cite{Koz,Zha},
  the number of ways that yields the state with the total angular
momentum J by summing 2N elementary spins $1/2$ is given by the formula:

\begin{equation}
\Gamma_N(J)=C_{2N}^{N-J}-C_{2N}^{N-J-1} \hskip5mm J=1,...,N.
\end{equation}

Then, (number of states with given numbers $S,J_1,L_1,J_2,L_2$) =
$\Gamma_{N_1}(J_1)\Gamma_{N_2}(J_2)$.
  Now, the Hamiltonian
depends on total angular momenta of the two groups of nuclei and, in the above
representation, can be broken down into blocks with specified values of the
total angular momenta ($J_1$ and $J_2$) of these groups.
  The fact that Hamiltonian (3)
has no matrix elements between the functions with different values of
 $\alpha$ can be
proven in the same way as in \cite{Koz}.
  The initial nuclear state (initial nuclear density
matrix) is assumed to correspond to infinite temperature. In this case, the
probability of each of the nuclear state is the same and equal to 1/(total
number of nuclear states) = 1/$2^{2(N_1+N_2)}$.
  Note that (dimensions of the block with given $J_1$ and $J_2$) =
(number of possible projections of electron spin)$\times$(number of possible
projections of $J_1$)$\times$(number of possible projections of $J_2$)=
$2(2J_1+1)(2J_2+1)$. These blocks are
independent and, hence, the dynamic equations for the density matrix can be also
broken down into blocks. The greatest possible nuclear polarization for the
considered case of two groups of equivalent nuclei is obtained for full nuclear
polarization in each of the blocks. In this case, the nuclear density matrix
 $\rho_n$
has, in such a block, only one nonzero diagonal element corresponding to
projection of the total nuclear spin equal to $J_1+J_2$:

  \begin{equation}
  \langle J_1,L_1=J_1,J_2,L_2=J_2|\rho_n|J_1,L_1=J_1,J_2,L_2=J_2\rangle=
  {(2J_1+1)(2J_2+1)\over 2^{2(N_1+N_2)}}
  \end{equation}

The number of such blocks in the nuclear density matrix equals
 $\Gamma_{N_1}(J_1)\Gamma_{N_2}(J_2)$. Projection of
the total nuclear angular momentum in this state is given by the expression

\begin{equation}
\langle J_z\rangle_{\hbox{max}}={1\over 2^{2(N_1+N_2)}}\sum_{J_1=0}^{N_1}\sum_{J_2=0}^{N_2}
\Gamma_{N_1}(J_1)\Gamma_{N_2}(J_2)(2J_1+1)(2J_2+1)(J_1+J_2)
\end{equation}

Taking into account the normalization condition \cite{Koz} for function (5)

\begin{equation}
{1\over 2^{2N}}\sum_{J=0}^{N}
\Gamma_{N}(J)(2J+1)=1,
\end{equation}

we have:

\begin{equation}
\langle J_z\rangle_{max}={1\over 2^{2N_1}}\sum_{J_1=0}^{N_1}
\Gamma_{N_1}(J_1)(2J_1+1)J_1+{1\over 2^{2N_2}}\sum_{J_2=0}^{N_2}
\Gamma_{N_2}(J_2)(2J_2+1)J_2=
\end{equation}
$$
=2N_1 P(N_1)+2N_2 P(N_2)
$$

Thus, for the maximum nuclear polarization  $P_2$, in the case
of two groups of equivalent nuclei, we obtain:

\begin{equation}
P_2={\langle J_z\rangle_{max}\over 2(N_1+N_2)}=\xi_1^2 P(N_1)+\xi_2^2 P(N_2)
=\xi_1^2 P(\xi_1^2 N)+\xi_2^2 P(\xi_1^2 N)
\end{equation}

where  $\xi_i^2\equiv N_i/(N_1+N_2)$ --
  is the fraction of nuclei of the i-th group and
  $N=N_1+N_2$ is the total
number of {\it pairs}
  of the nuclei.

  For arbitrary number of groups of the equivalent
nuclei $n$, we can obtain, in a similar way, the following expression for the
limiting polarization:

\begin{equation}
P_n=\sum_{i=1}^n \xi^2_i P(\xi_i^2N),\hskip10mm \sum_{i=1}^n\xi_i^2=1
\end{equation}

In the above treatment, the number of nuclei in each group was supposed to be
even. Intuitively, it seems plausible that this assumption is not essential
provided that the number of nuclei in a group  $\xi_i^2 N$ is much greater than unity. In
this case, we can use the approximate formula (2) for $P(N)$ and obtain:

 \begin{equation}
 P_n\approx {1\over\sqrt{\pi N}}\sum_{i=1}^n|\xi_i|
 \end{equation}

For a given number of groups of equivalent nuclei, the greatest nuclear
polarization is achieved when all the groups have the same number of nuclei,
i.e., $\xi_i^2=1/n$. In this case, the limiting polarization, for a given total number of
pairs of nuclei, exceeds by a factor of $\sqrt{n}$
 that of the box-model  $1/\sqrt{\pi N}$. Thus, if,
for instance, the real wave function of the electron in the quantum dot is
presented by 10 levels, the limiting polarization will increase, in this case,
not more than by a factor of $\sqrt{10}$.
 In the above calculation, the greatest nuclear
polarization arises when the number of groups of equivalent
 nuclei appears to be
equal to the number of nuclear pairs $N$. In this case, Eq. (12)
is not precise because the number of nuclei in a group is not
large (two nuclei in each). However the limiting polarization can
be now obtained using Eq. (11):

\begin{equation}
P_N={3\over 8}
\end{equation}
which is close to $1/\sqrt{\pi}$ obtained by Eq. (12). From the
above treatment we see, that the model in which the number of
groups of equivalent nuclei is only by a factor of 2 smaller than
their total amount, cannot describe the 100$\%$ nuclear
polarization.

\section{Discussion}

        As was already mentioned above, for the
 quantum dot consisting of $10^4$
nuclei, the box-model (with the number of groups of equivalent nuclei equal to
unity) yields the limiting nuclear polarization $\sim 1\%$.
  As follows from Eq. (12),
in the model compatible with the nuclear polarization of 50$\%$ obtained in
\cite{Fifty},
one has to take into account at least  $n=50^2=2500$
 groups of equivalent nuclei. It looks
hopeless to obtain exact solution for this model or to analyze it numerically.
The only possible approach to these problems can be based on uncontrollable
simplifications of its mathematical solution that imply more or less convincing
intuitive considerations.

Interpretation of a real experiments using such approaches appears
to be vulnerable in two points: (i) inaccuracy of the model itself
(insufficiently large number of groups of equivalent nuclei,
neglecting dipole-dipole interaction between nuclear spins, etc.),
(ii) uncontrollable errors related to simplification of
mathematical solution of the model problem. On the other hand, if
the experimentalist has some grounds to belive that the nuclear
polarization, in his experiments, does not
 exceed the limiting value $P_1$
for the box-model, then the use of this model (with only one of the two
vulnerable points listed above) for the
 interpretation of these experiments may
be quite efficient. The models that are more
  precise than the box-model include
mechanisms providing nuclear polarization higher than $P_1$.
  However, if the nuclear
polarization, in the experimental run, does not exceed $P_1$
  (e.g., under
sign-alternating optical orientation or under conditions of strong nuclear
relaxation), then these mechanisms can be considered as non-efficient
(sign-alternating orientation) or suppressed (strong nuclear relaxation), and
the box model can be used to interpret the experimental data.

Keeping in mind all the aforesaid, let us discuss, in a more
general sense, applicability of Hamiltonian (1) for description of
the experiments on spin dynamics of quantum dots. Within what time
intervals can Hamiltonian (1) be used for description of the
electron-nuclear spin dynamics in a quantum dot? How many groups
of equivalent nuclei should be chosen? The experiments on spin
dynamics of a quantum dot show that, for the repetition rate of
the pump pulses of about 100 MHz, the electron polarization
reaches its steady-state value
 ($\sim$ 1/3 in zero
magnetic field) for the time much shorter than the pulse repetition period. The
estimates show that, within this time scale, the
 internuclear dipole-dipole spin
interaction and the electron-nuclear dipole interaction \cite{Mer}, neglected in
Hamiltonian (1), are of no importance.
  The state of nuclear subsystem cannot
substantially change during one period of optical orientation,
because the angular momentum transferred to the nuclear subsystem
for this time interval cannot exceed 1. For this reason,
application of Hamiltonian (1) to calculation of time dynamics of
the electron spin between the pump pulses seems justified.
However, in this case, along with initial value of the electron
spin, one has to specify initial state of the nuclear subsystem.
Consider now, in a qualitative way, the nuclear dynamics. If we
assume that the nuclear system acquires from electron, with each
pump pulse, the angular momentum 1,
  then the quantum dot consisted of
$10^4$ nuclei with spins 1/2 will reach the degree
  of nuclear polarization of 1$\%$ not
earlier than after 100 pulses. Such a nuclear polarization is compatible with
the box-model. This is why, during, at least, the first 100 pulses, the
electron-nuclear spin dynamics calculated in
 the framework of the box-model will
hardly strongly differ from that calculated using the model with larger number
of groups of equivalent nuclei. Moreover, the studies of exact solution of the
box-model allow one to point out the reasons why, in fact, more than 100 pulses
are needed to achieve the nuclear polarization of $\sim 1\%$.
  The point is that, under
periodic orientation of the electron spin in the quantum dot, a regularity
arises in the nuclear density matrix \cite{Koz}
  leading to quasi-periodic dynamics of
the electron spin. The quasi-periodicity reveals itself in the fact that, just
before the arrival of the next pump pulse, the electron polarization in the
quantum dot increases to the value close to that just after the previous pulse.
As a result, the rate of the angular momentum transfer to the nuclear system
substantially decreases and the nuclear polarization process slows down. The
above arguments provide grounds to conclude that the electron-nuclear spin
dynamics of the box-model will not strongly differ from that of the model with
larger number of the groups of equivalent nuclei within the time interval
essentially exceeding 100 periods of optical orientation.

In summary, we can conclude that: (i) The initial stage of
the electron-nuclear spin dynamics (when the nuclear polarization does not
exceed $P_1$), can be described in a plausible way using the
box-model. This stage corresponds to a few first hundreds of the pump pulses.
(ii) The refinement of the box-model aimed at description of the nuclear
polarization exceeding $P_1$ by increasing the number of groups of equivalent
nuclei $n$ is characterized, on the one hand, by a slow
 (square-root) increase of
the limiting polarization  $P_n\sim
 \sqrt{n}P_1$ and, on the other, by impossibility of the exact
mathematical analysis.



\end{document}